\documentclass[10pt]{article}

\usepackage{amsmath}
\usepackage{amssymb}

\usepackage{graphicx}

\usepackage{cite}

\usepackage{color} 

\usepackage[labelfont=bf,labelsep=period,justification=raggedright]{caption}

\makeatletter
\renewcommand{\@biblabel}[1]{\quad#1.}
\makeatother

\date{}

\newcommand{\bec}[1] {\begin{equation}\label{#1} }
\newcommand{\eec} {\end{equation} }
\newcommand{\beq}{\begin{equation} }
\newcommand{\eeq}{\end{equation}}
\newcommand{\bea}{\begin{eqnarray}}
\newcommand{\eea}{\end{eqnarray}}

\begin{document}

\begin{flushleft}
{\Large
\textbf{A minimal model of signaling network elucidates cell-to-cell stochastic variability in apoptosis}
}
\\

Subhadip Raychaudhuri
\\

Department of Biomedical Engineering, Biophysics Graduate Group, Graduate Group in Immunology, Graduate Group in Applied Mathematics, University of California, Davis, 95616 USA Tel: 530 754 6716

$\ast$ E-mail: raychaudhuri@ucdavis.edu
\end{flushleft}

\section*{Abstract}
Background: Signaling networks are designed to sense an environmental stimulus and adapt to it. We propose and study a minimal model of signaling network that can sense and respond to external stimuli of varying strength in an adaptive manner. The structure of this minimal network is derived based on some simple assumptions on its differential response to external stimuli. 

Methodology: We employ stochastic differential equations and probability distributions obtained from stochastic simulations to characterize differential signaling response in our minimal network model. Gillespie's stochastic simulation algorithm (SSA) is used in this study. 

Conclusions/Significance: We show that the proposed minimal signaling network displays two distinct types of response as the strength of the stimulus is decreased. The signaling network has a deterministic part that undergoes rapid activation by a strong stimulus in which case cell-to-cell fluctuations can be ignored. As the strength of the stimulus decreases, the stochastic part of the network begins dominating the signaling response where slow activation is observed with characteristic large cell-to-cell stochastic variability. Interestingly, this proposed stochastic signaling network can capture some of the essential signaling behaviors of a complex apoptotic cell death signaling network that has been studied through experiments and large-scale computer simulations. Thus we claim that the proposed signaling network is an appropriate minimal model of apoptosis signaling. Elucidating the fundamental design principles of complex cellular signaling pathways such as apoptosis signaling remains a challenging task. We demonstrate how our proposed minimal model can help elucidate the effect of a specific apoptotic inhibitor Bcl-2 on apoptotic signaling in a cell-type independent manner. We also discuss the implications of our study in elucidating the adaptive strategy of cell death signaling pathways. 


\section*{Introduction}

Cellular signaling networks are designed to sense an environmental stimulus and respond in a strength dependent manner. In this manuscript, we develop and study a minimal model of a signaling network that can respond to an external stimulus in a manner such that the activation is fast under a strong stimulus but slow if the stimulus is weak. We derive the minimal network by assuming that the cell-to-cell variability dominates the slow signaling activation, under weak stimuli, in order to adapt to a fluctuating environment. In such a scenario, population average over many cells cannot capture cell-to-cell variability in signaling. We employ stochastic differential equations and stochastic simulations to study the signaling response in our proposed minimal signaling network. We carry out a sensitivity analysis of the minimal model with respect to parameter variations that also provides simple quantitative relations connecting different parameter values. We further use probability distributions of signaling molecules to characterize differential signaling response in the minimal network and such distributions show very distinct types of behavior depending on the strength of the stimulus. Specifically, for the case of a weak stimulus, a characteristic bimodal distribution is obtained for the activation of a downstream signaling molecule indicating large cell-to-cell fluctuations. 

Interestingly, the results from our minimal stochastic signaling model capture the essential stochastic signaling behavior observed in simulations of complex apoptotic cell death signaling pathways \cite{raychaudhuri1}. Details of the apoptotic signaling response vary depending on the cell type under consideration and also on the type of apoptotic stimulus applied \cite{albeck, spencer, chang, goldkorn, raychaudhuri2, skommer}. Our developed minimal signaling network demonstrates that large cell-to-cell stochastic variability increases as the strength of the stimulus is decreased, a feature also observed in large scale simulations and experiments of apoptosis, irrespective of cell types and the stimulus types used in those studies \cite{raychaudhuri1, albeck, skommer}. Hence, the minimal signaling network developed here captures cell-type independent features of apoptosis signaling and thus can serve as a general signaling model of apoptosis. We also discuss how the study of such a minimal network can provide crucial insights into the process of biological evolution of apoptosis signaling pathways.

\section*{Results}

\subsection*{The minimal signaling network}

We designed our proposed minimal signaling network to respond to an external stimulus in a manner dependent on the strength of the stimulus.  
\newline (i) Under a strong stimulus, activation of the signaling network is fast and there is no need for cell-to-cell stochastic fluctuations; 
\newline (ii) whereas under a weak stimulus, activation of  the signaling network is slow with large cell-to-cell variability so that it can adapt to a fluctuating environment. 

We attempted to find a minimal set of signaling reactions that can achieve the above response, with the nature of signaling reactions being that of irreversible chemical reactions of the type $X \rightarrow X^{*}$. 

Rapid deterministic activation under a strong stimulus can be achieved through a one-step direct activation of downstream signaling molecules (Figure 1a (i)). However, this one-step pathway would generate a slow deterministic activation of the signaling network under weak stimuli. Hence, for the case of a weak stimulus, we need a minimum of two separate rate constants to generate slow activation with large cell-to-cell variability. First, a low-rate constant step is required to generate slow stochastic activation of signaling molecules as the low kinetic rate constant will lead to a low probability of signaling activation with enhanced stochastic effects. However, one slow step cannot generate large cell-to-cell stochastic fluctuations and therefore it is essential to add a second fast step. This fast reaction will activate the signaling network rapidly once a few molecules have been activated by the low-probability event of a slow step (Figure 1a (ii)).  This slow-fast combination will presumably lead to the slow stochastic response in a manner such that large cell-to-cell variability with all-or-none type activation of the signaling network is achieved. The same external stimulus activates both pathways of the minimal network. Hence, we need to combine the two pathways in such a manner that, under a strong stimulus the one-step deterministic (denoted by type A) pathway wins, and,s under a weak stimulus the stochastic (denoted by type B) pathway dominates. This is achieved by adding a fast step to the second stochastic pathway that connects the second pathway to the external stimulus before the slow-fast steps. The rate constant for that fast step is chosen to be greater than that of the deterministic reaction of the first pathway to ensure that the stochastic pathway dominates under weak stimuli.The final downstream signaling molecule in the two pathways are considered to be the same molecule so that the two pathways are joined in a loop structure (Figure 1b). We used a combination of stochastic differential equations and stochastic simulations to study the behavior of the signaling network as the strength of an external stimulus is varied.

\subsection*{Signaling in the minimal network: stochastic differential equations}

We derive a stochastic differential equation for the one-step irreversible reaction $X \rightarrow X^{*}$ by equating the corresponding Fokker-Planck equation with the continuum limit of the Master equation for that activation reaction \cite{gardiner}. The stochastic equation is given by 

\beq\label{stochastic}
dx  \approx -k x dt  -  \sqrt{kx} dW
\eeq

$x$ denotes the number of molecules of the signaling species $X$ that are activated to $X^{*}$. $k$ is the reaction rate constant and $dW$ denotes the differential white noise ($W(t)$ is a Brownian process).  The equation that governs the number of activated species is given by $dx^{*} = -dx = k x dt +  \sqrt{kx} dW$, which shows that the total number $ x  +  x^{*} $ of a signaling species remains conserved. Eq. ~(\ref{stochastic}) is a nonlinear stochastic differential equation (SDE) with interesting structure of the noise term: for a low rate constant ($k \ll 1$), the noise term may not vanish even in the presence of a large number of molecules, which is in contrast to earlier considerations where stochastic effects were thought to be important only when a small number of signaling molecules were involved \cite{arkin, fedoroff}. We are now able to write the set of stochastic differential equations that describes all four reactions in type A and type B pathways:

\bea\label{TypeA&B}
  && dx_1 = -k_1 x_1 dt -  \sqrt{k_1 x_1} dW 
,\nonumber \\
 &&  dx_2 = - k_2 x_2 dt -  \sqrt{k_2 x_2} dW
,\nonumber \\
  && dx_3 =  -k_{31} x_{3} dt -  \sqrt{k_{31} x_{3}}dW  - k_{32} x_{3} dt -  \sqrt{k_{32} x_{3}} dW \nonumber \\
\eea

The number of molecules of the activated species  corresponding to $x_{i}$ is denoted by $x_{i}^{*}$ and obeys the equation $dx_{i}^{*} = -dx_{i}$.  The reaction rates in Eq.~(\ref{TypeA&B})  are given by $k_{1} = k_{1}^{0}  x_{0}, k_{2} = k_{2}^{0}  x_{1}^{*},k_{31} = k_{31}^{0}  x_{0}$ and $k_{32} = k_{32}^{0}  x_{2}^{*}$. Thus the reaction rates depend on the numbers of activated signaling molecules from the previous reaction step. $x_{0}$ is taken as the external stimulus that activates the signaling network under consideration. We denote the initial concentrations of all the inactivated signaling species by $x_{1}(0) = x_{1}^{0}$, and $x_{2}(0) = x_{2}^{0}$, $x_{3}(0) = x_{3}^{0}$, and the initial number of activated molecules for all three species are set to zero. Hence, the total number of molecules for the $i$ th signaling species is taken to be $x_{i}^{0}$. Same variables are used in our Monte Carlo study of the signaling network. We first consider the solutions of the above set of equations for the type A and type B pathways separately and then study their combined behavior. 

{\it (a) Deterministic signaling through type A pathway:} This is the one step signaling reaction described by the $dx_{3}$ reaction in Eq. ~ (\ref{TypeA&B}) with $k_{32}$ for the type B pathway being set to zero. This equation has the generic form of the Eq. ~(\ref{stochastic}): $dx = -kx dt - \sqrt{kx} dW$  and is amenable to analytical solution. We make a change of variable $z = \sqrt{kx}$. The solution to the equation in terms of our new variable $z$ is given as 
 
\beq\label{TypeB_solution}
  z(t) = z(0) e^{-\frac{k}{2} t} - \frac{k}{2}  \int_{0}^{t} e^{- \frac{k}{2} (t-s)}  dW    
\eeq

$s$ is a variable that is integrated over and $dW$ is a stochastic differential ($W(s)$ is a Brownian process). The number of signaling molecules of $x_{3}$ is calculated from the above solution using $x_{3} = z^{2}/k$. The number of activated molecules is then estimated simply by using $x_{3}^{*} = x_{3}^{0} - x_{3}$.  The solution, using parameter values of $k = 1$ and $x_{3}^{0}=100$, is plotted in Figure 2a. The rate constant $k$ sets the time-scale of deterministic activation.   

{\it (b) Stochastic signaling through type B pathway:} Here we use the three steps signaling reactions described by the equation set ~(\ref{TypeA&B})  with $k_{31}$ for the type A pathway set to zero in the last equation. The signaling reactions in each step again have the generic form of Eq. ~(\ref{stochastic}) which is amenable to an analytical solution. However, such a scheme with step-by-step solutions where the solution in one step is fed into the reaction constant of the next step quickly leads to complicated expressions that are not easy to analyze. Hence we numerically solved the set of stochastic differential equations given for the type B pathway. In Figure 2b, we show the activation of the end point of the signaling network, namely the activated $x_{3}$ molecules, as obtained from the numerical solution of the type B reactions. Clearly, in the type B signaling, the average behavior does not capture cell-to-cell stochastic fluctuations. We also probe the robustness of such a behavior in terms of parameter variations. The upper bound of the rate constant for the intermediate step $k_{2}^{0}$ is constrained by two factors: (a) it has to be small ($k_{2}x_{2} < 1$)  to generate stochastic fluctuations in the type B pathway, and (b) it also has to be significantly smaller than the rate constant for the final activation step ($k_{32}^{0} \gg k_{2}^{0}$) to generate all-or-none type activation with large cell-to-cell variability. Even though there is no lower bound needed to be defined for the constant $k_{2}^{0}$, it sets the time-scale of the signaling activation. In addition, lower the value of $k_{2}^{0}$, lower the amount of stimulus needed for the type B pathway to get activated (or equivalently longer the time to switch to the type B pathway). Hence the slowness of the stochastic pathway under weak stimuli also determines the relative dominance between type A and type B pathways.   

{\it (c) Transition from deterministic to stochastic signaling in a combined network of type A and type B pathways:} By comparing the two coefficients in the equation for the $x_{3}$ activation, we can derive a condition for which the deterministic type A pathway will dominate over the stochastic type B pathway (over the entire time-course of activation). The relevant coefficient for the type A pathway is $\sim k_{31} = k_{31}^{0}x_{0}$ and the same for the type B pathway is $ \sim k_{32} = k_{32}^{0}x_{2}^{*}$. We approximate the intermediate reaction step of the type B pathway by the stochastic activation term such that $dx_{2} \approx - \sqrt{k_{2}x_{2}} dW \approx - \sqrt{k_{2}^{0}x_{1}^{*} x_{2}^{0}} dW$. Activation of the first step of the type B pathway occurs in a rapid manner and we could approximate $x_{1}^{*} \approx x_{1}^{0}$, and $k_{32}^{0}x_{2}^{*}  \sim k_{32}^{0} \sqrt{k_{2}^{0}x_{1}^{0}x_{2}^{0}}$. Thus comparing the coefficients $k_{31}^{0}x_{0} \sim k_{32}^{0} \sqrt{k_{2}^{0}x_{1}^{0}x_{2}^{0}}$, we derive a condition for which the deterministic type A pathway will dominate over the stochastic type B pathway: $x_{0} > \frac{k_{32}^{0}  \sqrt{k_{2}^{0} x_{1}^{0}x_{2}^{0}}}{k_{31}^{0}}$ (Supplemental Figure 1). Hence, below a certain threshold value of the stimulus $x_{0}$, the stochastic type B pathway plays a role and above that value the deterministic type A signaling leads to rapid activation of $x_{3}$. Note, the observed transition from deterministic to stochastic behavior in signaling, depending on the strength of an external stimulus, confirms that the signaling network under consideration has a suitable minimal structure. However, several simplifying assumptions were made in the calculations based on stochastic differential equations. We use Monte Carlo simulations for a detailed study and also to corroborate the results obtained from our theoretical analysis.

\subsection*{Signaling in the minimal network: Monte Carlo simulations}

We use GillespieÕs Monte Carlo sampling technique \cite{gillespie} to simulate the Master equations that describe the set of chemical reactions given in the equation Eq. ~(\ref{TypeA&B}). The initial condition is set as $x_{1}^{0} = 100, x_{2}^{0} = 50, x_{3}^{0} = 50$. We consider activation of the downstream signaling molecule $x_{3}^{*}$ (for $k_{31}^{0} \ll k_{1}^{0}$) as we vary the external stimulus $x_{0}$.       

{\it Slow stochastic activation under a weak stimulus:} For a weak external stimulus  $(x_{0} < 10)$, large cell-to-cell fluctuations dominate the signaling behavior, where $x_{3}$ activation at the single cell level is rapid compared to the time-scale range for $x_{3}$ activation for all cells (Figure 3: {\it left panel}). The plot was generated for $x_{0} = 1$ and with reaction constants set to: $k_{31}^{0} = 10^{-5}$ (for the type A pathway), and $k_{1}^{0} = 1.0$, $k_{2}^{0} = 10^{-7}$, and $k_{32}^{0} = 1.0$ (for the type B pathway) in suitable units. We derive a condition for which the signaling will be entirely dominated by the type B pathway by comparing $k_{31}^{0}x_{0}x_{3}^{0} \sim k_{2}^{0}x_{1}^{0}x_{2}^{0}$. For the given parameter values, we obtain $x_{0} \sim 1$ (or less) for pure type B signaling, which agrees well with our numerical simulation results (Figure 3). Decreasing $k_{2}^{0}$ would demand increasingly lower $x_{0}$ for the type B pathway to dominate at early times of signaling (Supplemental Figure 1). The small rate constant in the intermediate step of the type B pathway leads to slow activation of the pathway where different cells initiate activation at very different time-points and then a subsequent fast rate constant leads to a spike in $x_{3}$ activation. We call this phenomenon all-or-none type activation with large cell-to-cell variability. Decreasing stimulus slows down the downstream signaling and increases cell-to-cell stochastic variability. We can estimate the variance of $x_{3}$ activation at the single cell level using a sharp-rising approximation scheme for $x_{3}$ activation (ignoring the spread along the time axis for a single cell). Suppose, $m(t)$ is the number of cells, out of a population N cells, in which $x_{3}$ got activated within a given time period $t$. Assuming rapid activation of $x_{3}$, the average of $x_{3}$ is given by $<x_{3}> = x_{3}^{0}(1-m(t)/N)$ and the average of $x_{3}$ square is $<x_{3}^{2}> = (x_{3}^{0})^{2} (1-m(t)/N)$. The variance can then be estimated as $<x_{3}>^{2} - <x_{3}^{2}>  = (x_{3}^{0})^{2} (1-m(t)/N)(m(t)/N)$. Hence a simple scaling relation is obeyed by the Fano factor, {\it i.e.} variance($t$)/average($t$), $ = x_{3}^{0} (m(t)/N)$, which remains $ \gg 1$ and increases with time implying presence of large cell-to-cell stochastic fluctuations.  Calculation of the Fano factor from actual simulation data (such as in figure 3: {\it left panel}) also shows that Fano factor at a given time increases with increasing $x_{3}^{0}$ and remains larger than one as predicted by the scaling relation. However, the scaling relation is obtained only within a sharp-rising approximation and thus slightly deviates from the variance/average ratio obtained from simulations.   

The stochastic signaling observed in response to a weak external stimulus is essentially dominated by the type B pathway, as a similar signaling response is generated if the kinetic constant for the type A pathway is set equal to zero. This result is also consistent with our theoretical analyses of the minimal network that showed the presence of large stochastic fluctuations under a weak stimulus.  Stochastic signaling behavior through the type B pathway in response to a weak stimulus is maintained as long as the fast-slow-fast kinetics of signaling reactions is also maintained in the type B pathway (Supplemental Figure 1). For a stimulus of intermediate strength ($x_{0} \simeq 10-1000$), a mixed signaling behavior is observed with a sudden change from type A to type B at the level of single cells (Figure 3: {\it middle panel}). {\it Also note, for pure type B signaling (when the kinetic constant for type A is set to zero), presence of a large number of molecules of an initial stimulus ($x_{0} \gg 1$) does not change the observed stochastic signaling behavior.}  

{\it Rapid deterministic activation under a strong stimulus:} Under a strong external stimulus, i.e. for large values of $x_{0}$ ($> 1000$), the type A pathway dominates the signaling behavior (Figure 3: {\it right panel}), as also observed in the theoretical study using stochastic equations. The plot was generated for $x_{0} = 1000$ and with reaction constants set to: $k_{31}^{0} = 10^{-5}$ (for the type A pathway), and $k_{1}^{0} = 1.0$, $k_{2}^{0} = 10^{-7}$, and $k_{32}^{0} = 1.0$ (for the type B pathway) in suitable units. For these given parameter values, our theoretical estimate of the threshold stimulus $x_{0} > \frac{k_{32}^{0}  \sqrt{k_{2}^{0} x_{1}^{0}x_{2}^{0}}}{k_{31}^{0}}$ results in $x_{0} > 1000$ (order of magnitude approximation), which matches well with our stochastic simulation results. Similar deterministic signaling is observed when the kinetic constant for the type B pathway is set to zero to assure a pure type A response. As mentioned earlier, characteristics of such type A signaling is small cell-to-cell fluctuations where the average (over a population of cells) behavior dominates the signaling.  As we decrease the strength of an external stimulus increased cell-to-cell stochastic fluctuations are observed with a gradual switch to the type B activation. 

{\it Probability distribution approach to characterize stochastic signaling:} 
Large ($ > 1$) variance/average ratio for weak stimuli indicates large stochastic variations in signaling. To characterize such stochastic fluctuations in the minimal network we determine the probability distribution of activated $x_{3}^{*}$. A histogram of activated $x_{3}^{*}$ is generated from many runs (i.e. many cells) of our simulations that defines the probability distribution of $x_{3}^{*}$. In the case of a strong stimulus $x_{0} = 1000$, i.e. when the type A pathway dominates the signaling, the probability distribution shows a gradual shift (along the x axis) over time.  Thus the average behavior is representative of the entire cell population and cell-to-cell stochastic fluctuations are not significant when compared to the average (Figure 4, top panel). The plot was generated for $x_{0} = 1000$ and with reaction constants set to: $k_{31}^{0} = 10^{-5}$ (for the type A pathway), and $k_{1}^{0} = 1.0$, $k_{2}^{0} = 10^{-7}$, and $k_{32}^{0} = 1.0$ (for the type B pathway) in suitable units. The initial condition is set as $x_{1}^{0} = 100, x_{2}^{0} = 50, x_{3}^{0} = 100$. On the contrary, if the stimulus is weak ($x_{0} = 1$), large cell-to-cell stochastic fluctuations lead to a bimodal distribution (Figure 4, bottom panel) for activated $x_{3}^{*}$ molecules where one peak gets taller while the other peak gets shorter. Such bimodal distribution arises because activation of $x_{3}^{*}$ in an individual cell is completed in a period of time that is orders of magnitude smaller than the time over which activation occurs for a large population of cells.  We define such signaling response as all-or-none type activation with large cell-to-cell variability. Hence, such a bimodal probability distribution can be considered as a characteristic of signaling through the type B pathway with large stochastic variations. For intermediate stimuli, the probability distribution shows characteristics of both type A and type B activation (Figure 4, middle panel).

\subsection*{The minimal network elucidates signaling mechanisms in apoptosis}

Apoptotic cell death signaling pathway is one of the most complex intracellular signaling networks and consists of a large number of components \cite{gomperts, watters}. Large number of signaling components, molecules, and complicated network structure all contribute to the enormous complexity of the apoptotic cell death signaling pathways, and thus mask the essential design principles of apoptosis signaling. However, one may want to know if there exists a minimal signaling network structure that can capture the essential qualitative features of the cell death signaling through the apoptotic pathways. 

Apoptotic cell death signaling is generally mediated by two distinct pathways that are connected in a unique loop structure through the final downstream activation of caspase-3 molecules \cite{watters}. Activation of the apoptotic cell death signaling pathway has been shown to have two disparate time scales: (a) fast (minutes) activation in some cases such as (\cite{scaffidi}), and (b) orders of magnitude slower ($>$ 10 hrs) activation under certain conditions \cite{goldkorn, scaffidi}. However, the presence of a specific signaling molecule and its concentrations are often cell-type specific \cite{porwit-macdonald}, calling into question the robustness of the results obtained from experiments studying apoptosis signaling. Our proposed minimal model of stochastic signaling network can capture the experimentally observed slow activation of the apoptotic signaling pathway in a cell-type independent and robust manner.  The slow intermediate step in the type B pathway of the minimal network captures the slow reaction of Bax activation or apoptosome formation (or both of those) in apoptosis signaling. The time-to-death and its cell-to-cell variability in apoptosis signaling is limited by stochastic Bax activation induced cytochrome c release and apoptosome formation induced caspase-9 activation \cite{raychaudhuri1, raychaudhuri2, skommer}. The relative contribution of those two slow stochastic events in apoptosis can vary significantly depending on the cell type and apoptoptic stimuli used. We could effectively simulate the effect of both slow steps in apoptosis, and their stochastic variations, by varying the rate constant of the intermediate slow step in the type B pathway of the minimal model (Figure 5). Thus the proposed minimal network with stochastic signaling behavior seems to be an appropriate minimal model for the apoptosis signaling where large cell-to-cell fluctuations cause slow cell death. This minimal network could also be derived from the full apoptotic signaling network by (i) grouping functionally redundant proteins and (ii) replacing a set of reactions by a single effective reaction with modified rate constants. However, the approach taken in this work has the advantage that prior knowledge of the apoptotic pathway is not needed, instead, a simple set of assumptions on the response of the signaling network is sufficient to generate the minimal network structure.     

Results of our minimal model indicate that a slow rate constant followed by a faster kinetics in our minimal model is enough to generate large cell-to-cell stochastic variations as observed in the large scale simulations,  as well as in single cell experiments of apoptosis signaling \cite{raychaudhuri1, albeck, spencer, raychaudhuri2, skommer}. Variations in the initial number of signaling molecules in the type B pathway will enhance the cell-to-cell stochastic variability, a feature also observed in apoptosis signaling \cite{skommer}. We have further connected our minimal model of stochastic signaling with a specific model of apoptosis inhibition mediated by anti-apoptotic Bcl-2 protein. The loop structure of tBid-Bcl-2-Bax along with the Bcl-2 level determine the time course of Bax2 activation and thus apoptotic activation through the mitochondrial pathway of apoptosis \cite{watters, skommer}. The rate constants for the tBid-Bcl-2, tBid-Bax and Bax-Bcl-2 reactions are not small and the number of molecules initially present are also large. However, inhibition by Bcl-2 molecules reduces the effective rate for Bax conversion to activated Bax and subsequent Bax-2 complex formation. In cancer cells, Bcl-2 inhibits over-expressed BH3 molecule such as Bid, which can directly activate Bax with a very small rate constant \cite{skommer,letai}, and thus through the Bid-Bcl-2-Bax loop stochastic Bax activation is generated. Such Bcl-2 inhibition of Bax activation dynamically generates stochastic variability in apoptosis even when the initial number of molecules present are not small \cite{arkin, fedoroff}. In our minimal model, we have an effective $X \rightarrow X^{*}$  reaction with a slow reaction rate constant which can be decreased to simulate an increase in Bcl-2 levels. Recent experimental and theoretical studies elucidated that increase in Bcl-2 levels increases time-to-death and its cell-to-cell variability in apoptosis signaling \cite{skommer}. For very high Bcl-2 levels, as observed in some cancer cells, apoptotic activation is strongly inhibited with rare stochastic activations of a few cells \cite{skommer}. In our minimal model, reduction in the rate constant of the slow intermediate step of stochastic pathway generates slower activation with increased cell-to-cell variability and thus captures the signaling behavior in apoptosis with over-expressed Bcl-2 levels (Figure 5).

\section*{Discussion}

In this article, we proposed a minimal signaling network that is capable of sensing an environmental stimulus and generating an appropriate signaling response in accordance with the strength of the stimulus. We derived this network structure starting with a simple set of assumptions related to the differential response of this signaling network to varying external stimuli. Specifically, under a weak stimulus, the minimal network is designed to use slow stochastic signaling to adapt to a fluctuating environment. We think the proposed minimal network can serve as a minimal signaling model for the complex cell death signaling pathway. Having large cell-to-cell stochastic fluctuations could be a strategy which cells use to respond to a weak environmental stimulus and thus diversify their options for adapting to a fluctuating environment. We hope our results can be tested using synthetic networks and will guide the engineering of biosynthetic mimics designed to sense the environment in the proposed manner. We also used a probability distribution based approach for the characterization of stochastic signaling that shows non-trivial bimodal distribution under weak stimuli. 

Large-scale complex signaling pathways, such as cell growth, cell death or immune response signaling pathways, were thought to behave in a deterministic fashion due to the presence of a large numbers of molecules \cite{arkin, fedoroff, oudenaarden, swain}. In this work, we show that stochastic fluctuations can persist even in the presence of a large number of molecules . This result becomes even more important in light of our recent simulations of the apoptotic cell death pathway, where stochastic signaling behavior is also observed. The study of minimal models prove the robustness (cell-type independence) of the results obtained from our Monte Carlo simulation of the apoptosis signaling network and thus can elucidate the essential design principles of apoptotic cell death signaling in normal and diseased cells. 

In addition, the proposed minimal signaling network can provide crucial insights into the adaptive evolutionary strategy of complex cell death signaling network. The CED3-CED4 death pathway found in C. Elegans resembles the type B pathway of the proposed minimal signaling network \cite{yuan}.  The large-scale type 2 apoptotic network could have evolved such complexity in response to the needs of increasingly complex and higher level species. Also note, the type 1 apoptotic pathway, which induces fast deterministic activation under strong apoptotic stimuli, first appeared in vertebrates concurrent with the first appearance of adaptive immunity. This type 1 apoptotic pathway resembles the deterministic type A pathway of our minimal network and could have first appeared in order to control adaptive immune response through rapid activation of the apoptotic programmed cell death. Thus the proposed minimal stochastic signaling network could be an evolutionarily conserved network motif and thus could serve as a general model for signaling during apoptosis. Such a minimal model based approach to elucidate design principles of cell signaling pathways is very general and can also be applied to study other large-scale signaling systems.


\section*{Acknowledgments}

S.R. thanks S.C. Das for reading of the manuscript and M. Djendinovic for assistance with preparation of figures.

\bibliography{template}




\section*{Figure Legends}

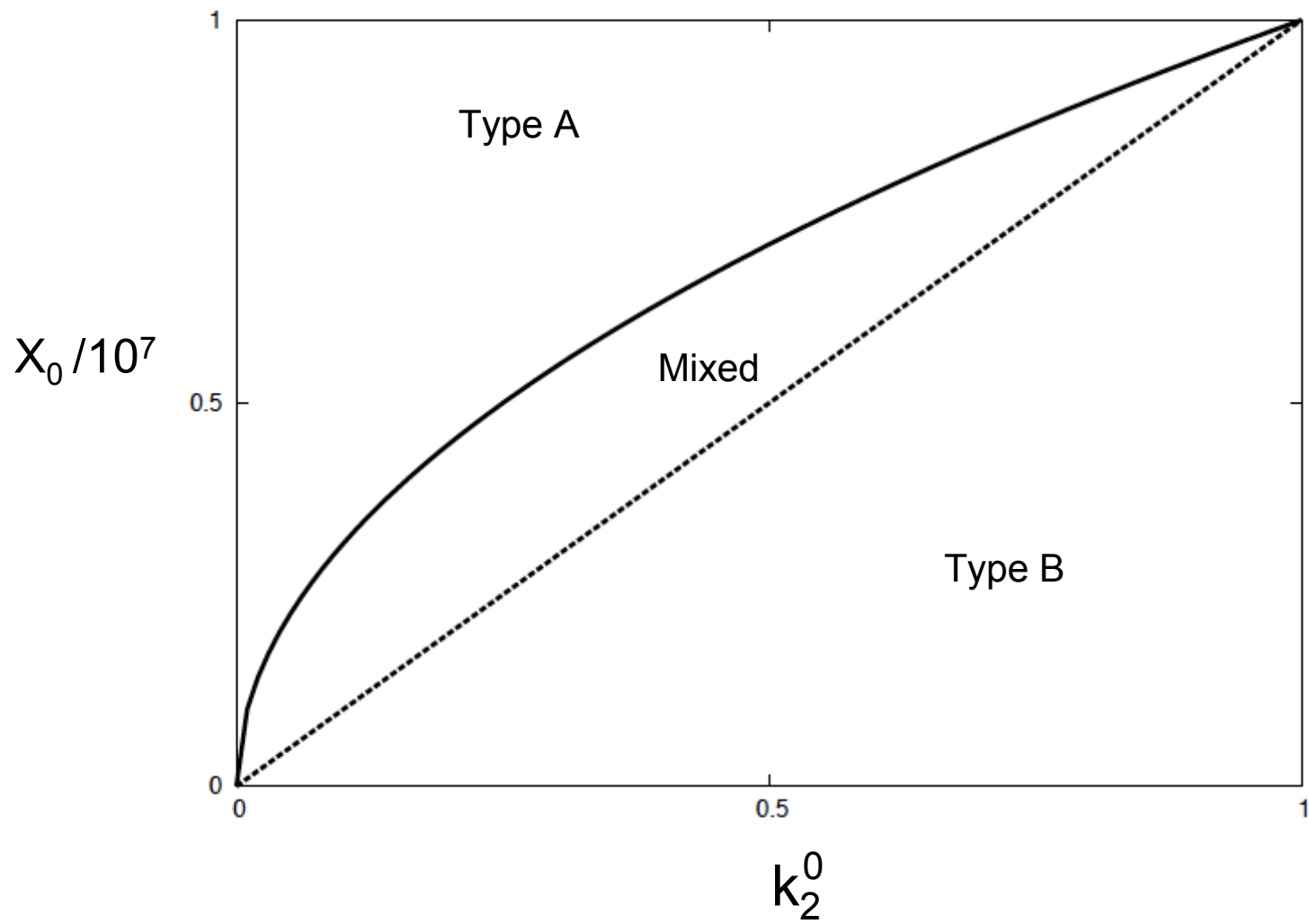
\begin{figure}[!ht]
\caption{
{\bf  Schematics of minimal signaling networks.} (a) Schematics of a minimal deterministic signaling network (i), and the same for a two-step signaling network with slow stochastic activation (ii). (b) Schematics of the minimal signaling network }
\end{figure}

\begin{figure}[!ht]
\caption{
{\bf  Differential signaling through type A and type B pathways.} (a) Plots obtained from the exact solution of the stochastic differential equation for type A signaling. (b) Plots obtained from the numerical solution of the stochastic differential equations for type B signaling. Each colored line corresponds to a solution for a single realization of the randomness {\it i.e.} a single cell (for both (a) and (b)). Note the disparate time-scales of signaling activation in (a) and (b). }
\end{figure}

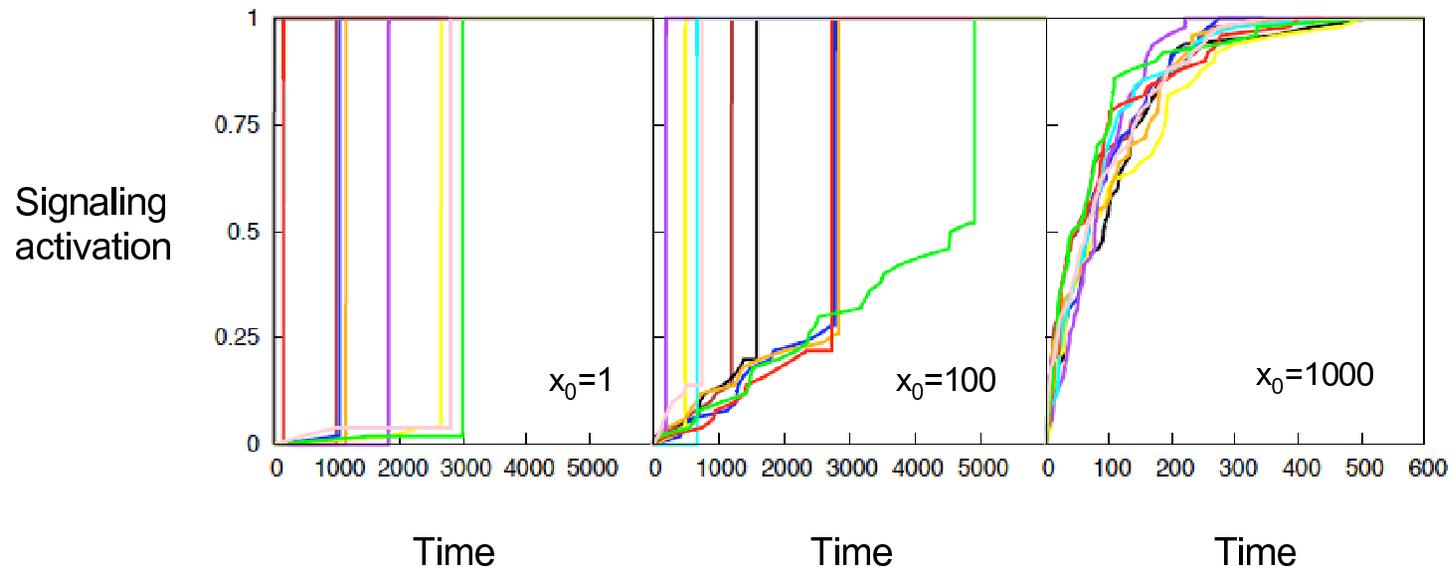
\begin{figure}[!ht]
\caption{
{\bf Switch from type B to type A signaling with increasing stimuli.}  Activation of $x_{3}$ molecules at the single cell level for weak: $x_{0} = 1$ (left), intermediate: $x_{0} = 10$ (middle), and strong: $x_{0} = 1000$ (right) stimuli. $x_{3}$ activation is normalized by its maximum. Different colors correspond to different individual cells.}
\end{figure}

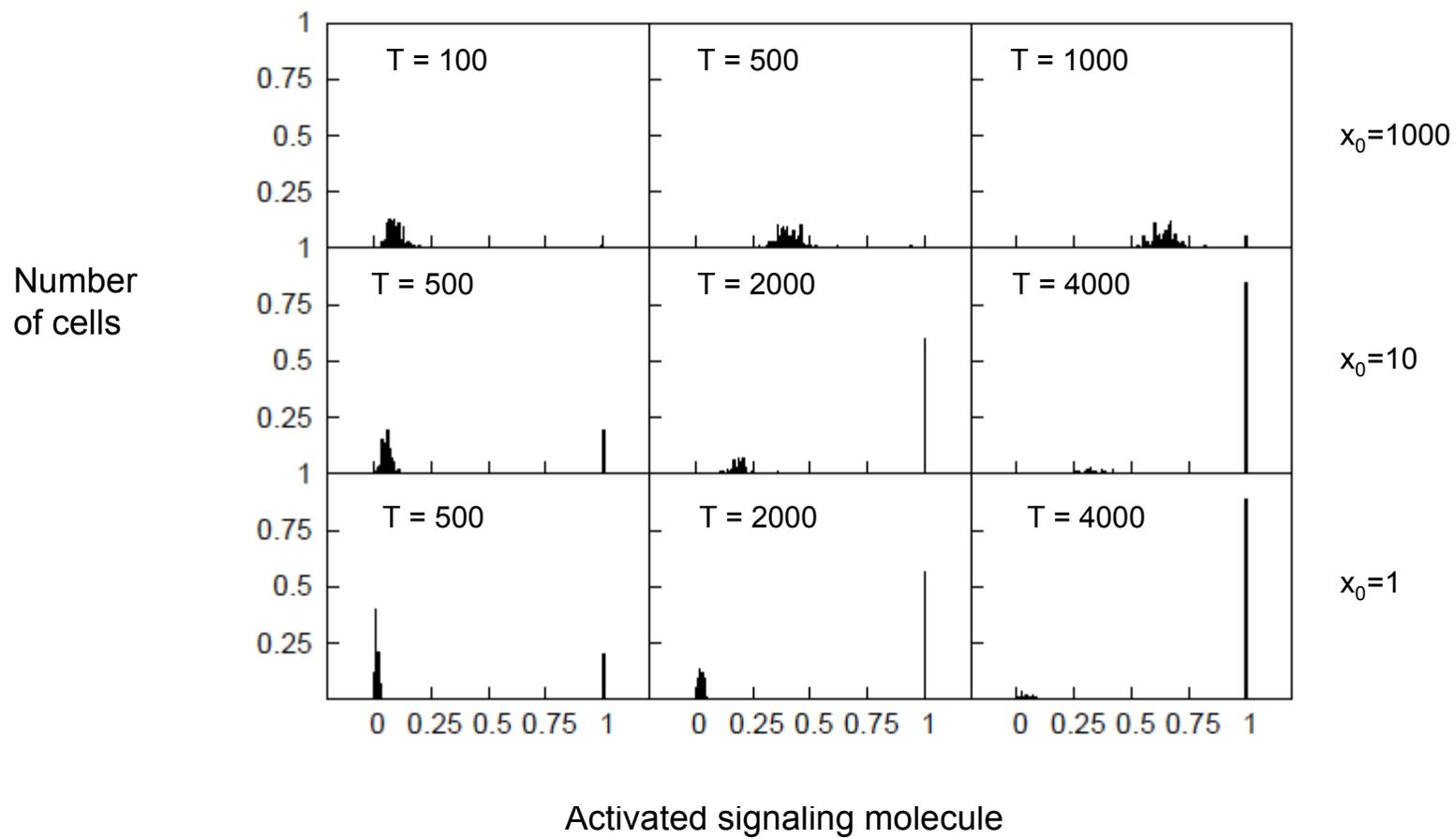
\begin{figure}[!ht]
\caption{
{\bf Probability distribution as a measure of cell-to-cell stochastic variability.} Probability distribution of activated $x_{3}$ molecules under strong $x_{0} = 1000$ (top) , intermediate $x_{0} = 10$ (middle), and weak $x_{0} = 1$ (bottom) stimuli. Histograms were obtained using single cell $X_{3}$ activation data from a population of 100 cells. For the strong stimulus (top) probability distributions are shown for three time points T=100, T=500, and T=1000. For the intermediate stimulus (middle), probability distributions are shown for three time points T=500, T=2000, and T=4000. For the weak stimulus (bottom), probability distributions are shown for three time points T=500, T=2000, and T=4000.  }
\end{figure}

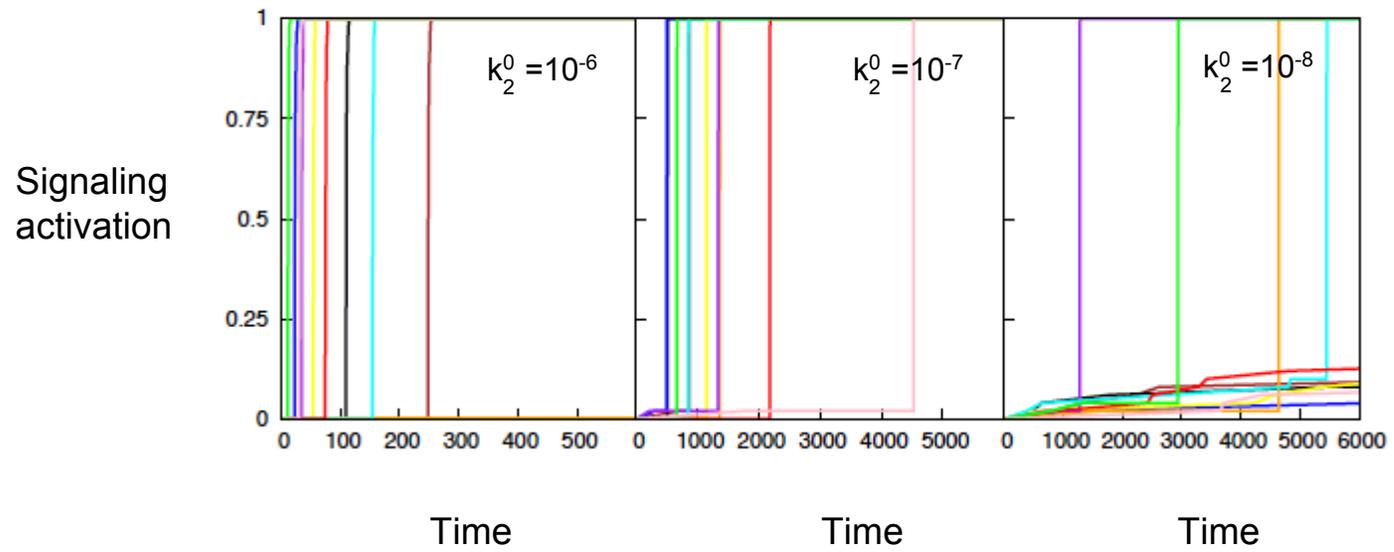
\begin{figure}[!ht]
\caption{
{\bf The slow step in the type B pathway modulates cell-to-cell stochastic variability.}  Activation of $x_{3}$ molecules at the single cell level for various values of the rate constant for the slow intermediate step (type B pathway): $10^{-6}$ (left) ,  $10^{-7}$ (middle), and $10^{-8}$ (right) stimuli. Different colors correspond to different individual cells.   }
\end{figure}


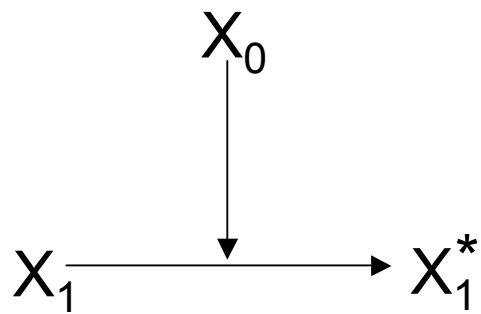
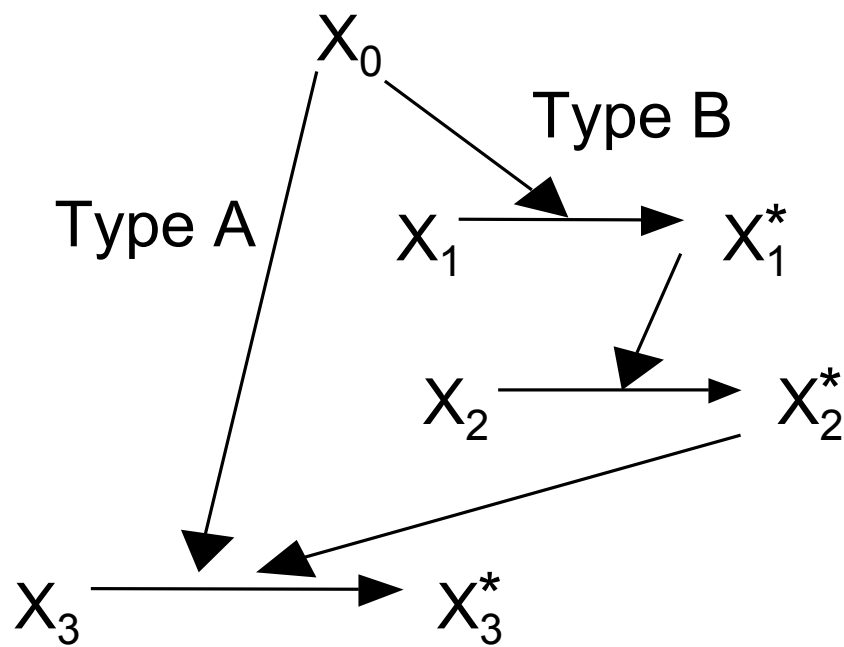
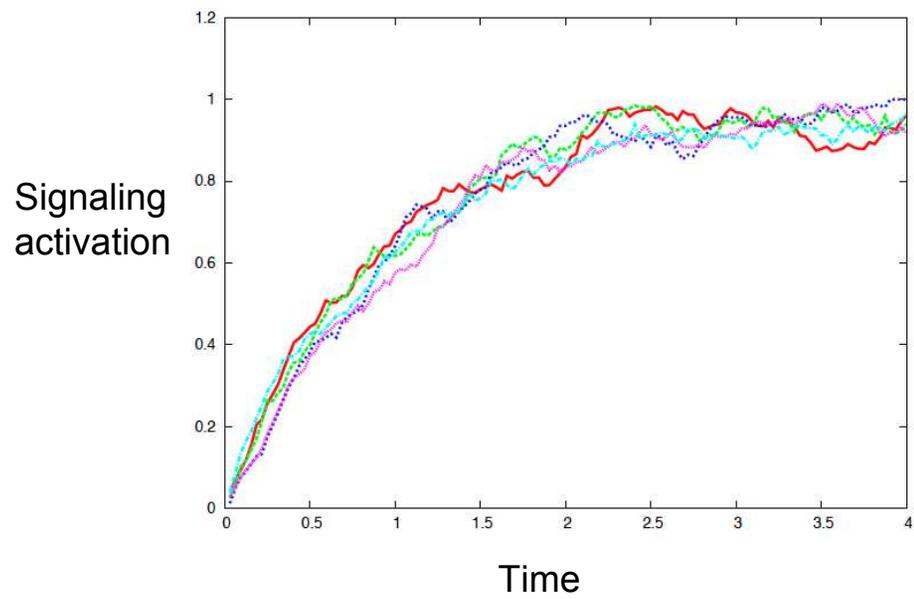
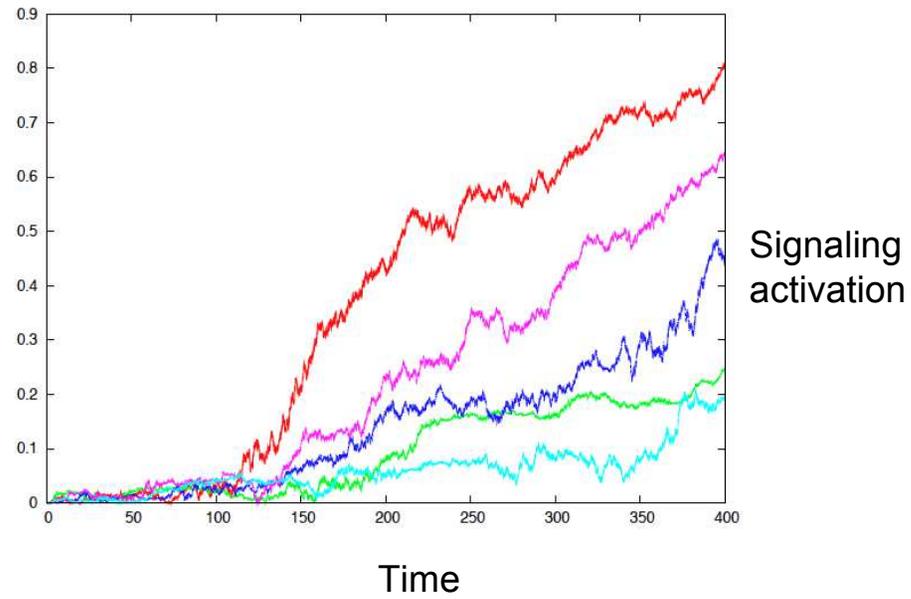
\end{document}